\documentclass[twocolumn,showpacs,preprintnumbers,amsmath,amssymb]{revtex4}
\usepackage{subfigure}
\usepackage{graphics}
\usepackage{multirow}
\usepackage{graphicx}
\usepackage{amssymb}
\usepackage{epstopdf}
\usepackage{footnote} 
\usepackage[T1]{fontenc}
\usepackage{bm}

\begin{document}


 \title{Quantifying social vs. antisocial behavior in email networks}
 		       
 \author{Luiz H. Gomes$^{1}$, Lu{\'i}s M. A. Bettencourt$^{2}$, Virgilio A. F. Almeida$^{1}$\\ 
                 Jussara M. Almeida$^{1}$ and Fernando D. O. Castro$^{1}$ \\
$^1$Computer Science Department \\
  Universidade Federal de Minas Gerais, Belo Horizonte, Brazil \\
$^2$Theoretical Division,  \\ 
           MS B284, Los Alamos National Laboratory, Los Alamos NM 87545, USA \\
}

\begin{abstract}
Email graphs have been used to illustrate general properties of 
social networks of communication and collaboration. However, increasingly, the majority  
of email traffic reflects opportunistic, rather than symbiotic social relations. 
Here we use e-mail data drawn from a large university to construct directed graphs of email 
exchange that quantify the differences between social and antisocial behaviors in networks of communication. 
We show that while structural characteristics typical of other social networks are shared 
to a large extent by the legitimate component they are not characteristic of antisocial traffic.
Interestingly, opportunistic patterns of behavior do create nontrivial graphs with certain general 
characteristics that we identify. To complement the graph analysis, which suffers from incomplete 
knowledge of users external to the domain, we study temporal patterns 
of communication to show that the dynamical properties of email traffic
can, in principle, distinguish different types of social relations. 
\end{abstract}

\maketitle
\date{\today}

\section{Introduction}
\label{sec:intro}

The fast pace of recent progress in the quantitative understanding 
of complex networks that mediate social interactions has been 
largely due to new ways of harvesting data, mainly by electronic 
means. For this reason graphs of email communication, where nodes 
represent email users and links denote messages exchanged between 
them, have become important example social networks.  The statistical mechanics of these networks makes possible a quantification of aspects of human social behavior and their comparison to the structure of interactions in other complex systems. 

A recent study \cite{Newman2003} has provided evidence for structural properties that are characteristic 
of social graphs, but not of other complex networks. These 
are a nontrivial clustering coefficient (network transitivity) 
and the presence of positive degree correlations (assortative mixing by degree)
 between adjacent nodes. Moreover, it has been suggested that social networks can be largely understood in terms of the organization of nodes into communities \cite{Newman2003,Guimera2003,Huberman2003,Arenas2004,
casado:opportunistic-hotnets05}, 
a feature that can explain, to some extent, the observed values for the 
clustering coefficient and degree correlations. 
This observation has indeed led to the interesting suggestion that email networks can be used to infer informal communities of practice within organizations \cite{Huberman2003}, as well as their hierarchical 
structure \cite{Guimera2003,Huberman2003,Guimera2004}, features 
that can in principle be useful for the efficient management of human collective behavior.   In fact, the nature of such hierarchies can be quantified  \cite{Guimera2003,Trusina2004}, and may be self-similar \cite{Guimera2003}. 

Beyond these characteristics that are, at least at the qualitative level, general to social networks 
there are features of email graphs that are specific. 
The most important property of email is the low cost \footnote{In energy, time and reputation of the sender} 
involved in delivering a message to a large group of recipients. This tends to make communication 
between any two nodes more indiscriminate, as email senders may easily send copies of a message 
to multiple parties that play no active role in the relationship between sender and recipient. As such, we may expect that networks of email may contain nodes with very high 
degree, and that degree distributions exhibit less severe or no practical constraints 
to their high degree tails. The result, as we show below, is that networks of email 
show no upper cutoff in their degree distributions, which are scale free with a small exponent 
$\alpha$, and degree correlations that may be atypical of other social networks. 

The ease with which messages can be distributed to many recipients is also at the root of most opportunistic behavior involving email. In fact, there has been growing interest in uncovering 
evidence of {\it antisocial} behavior in online networks. Recent work addresses topics such as 
uninhibited remarks, hostile flaming, non-conforming behavior,  group polarization, and spurious traffic~\cite{Wellman96,Gueorgi}. Email as a means of potential mass distribution is 
particularly associated with the dissemination of computer viruses as well as spam traffic \cite{spamhaus}, 
that flood the Internet with unwanted messages usually containing commercial propositions or, 
more recently, a variety of other scams. This behavior, which we call generically {\it antisocial}, displays 
different characteristics from other types of social relations for which social networks 
have been constructed and analyzed.

In all previous characterizations of email communications as networks, 
the problem that these networks also mediate antisocial relations has not been addressed. 
In order to attempt to eliminate such behaviors, as well as to deal with
 incomplete network reconstruction, authors have used several 
strategies such as restricting the analysis of email traffic to within the organization's 
domain \cite{Guimera2003,Huberman2003,Arenas2004,Guimera2004,Newman2002},
 taking into account only links that display communication in both 
directions \cite{Guimera2003,Arenas2004,Guimera2004}, eliminating nodes 
associated with very high message volumes \cite{Guimera2003,Arenas2004,Guimera2004},
and setting minimal message thresholds for a link to exist \cite{Huberman2003}.

Here we provide a more complete study of email networks by lifting 
most of these restrictions. Then email networks become directed, and 
the number of users and links in our dataset  is dominated by spam traffic.  What is conceptually interesting about spam email is that it nevertheless displays quantitative graph 
theoretical and dynamical characteristics that are nontrivial. Moreover, these characteristics reflect a certain type of antisocial behavior that can be quantitatively characterized and contrasted to the general properties of other social networks. 

The remaining of this manuscript is  organized as  follows. In section~\ref{sec:network} we give details about our data and the several networks of social, and antisocial behavior constructed. We then proceed to analyze them via standard network measures for which we expect antisocial behavior to differ from social. In section \ref{sec:temporal} we give an additional characterization of the temporal structure of time series of email and show that social and anti-social traffics differ in several characteristic ways. Finally we present our conclusions.

\section{Network inference and structural analysis}
\label{sec:network}

To construct networks of email communication we consider  the email traffic from a department of a large university. Email messages arriving at the departmental server are classified either as spam or legitimate by SpamAssassin, a standard and widely used filtering software \cite{spamassassin}. We construct four graphs representing different email networks. A {\it social network} is built from the legitimate (as classified by SpamAssassin) messages exchanged  between all users, including those external to the department that send/receive e-mails to/from internal users. Similarly, an {\it antisocial network} is built from the messages classified as spam, exchanged between all users. An {\it internal social network} is built by considering internal users exclusively involved in legitimate internal email communication. Finally, the internal spam traffic \footnote{Originating from and addressed to an internal user. These are usually the result of forged identifiers.} is used to build an {\it internal antisocial network}. In general these networks are directed. We note that messages exchanged through legitimate mailing lists, which also involve bulk email traffic, may exhibit antisocial characteristics. As in \cite{Eckmann2004}, aiming at minimizing the impact of such communication patterns in our analysis, we remove users who exchange emails with fifty or more other users from our internal social network.

Our four networks are built from a thirty-day log including 562664 messages,  
of which 270491 are spam. The set consists of 19504 internal and 259069 external users. Of these, 164998 external users are senders of spam, while that number is only 721 for those internal to the domain, most of them under fabricated identifiers.  Also note that the number of users in our log is orders of magnitude larger that those included in several previously analyzed datasets \cite{Ebel2002,Shetty2005}.

\begin{figure}[t!]
  \centering
\includegraphics[width=200pt]{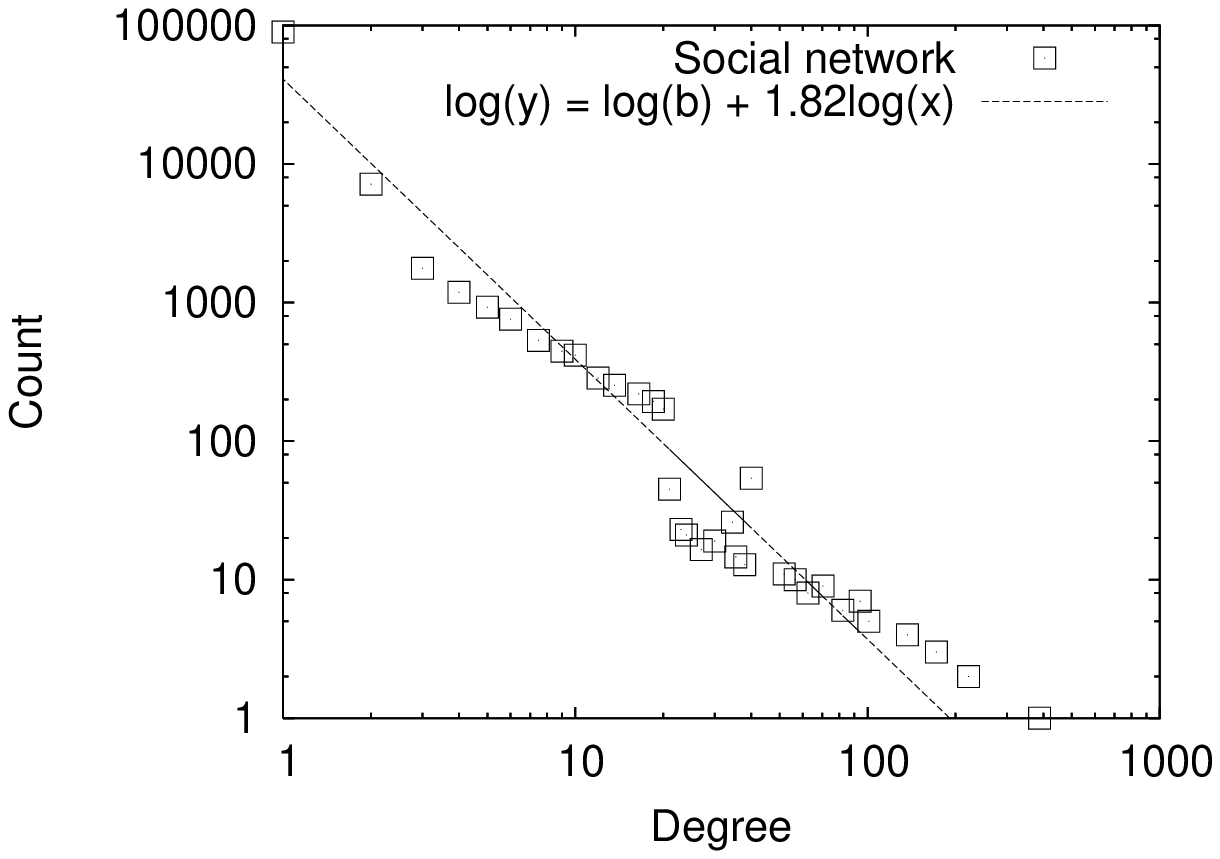}
\includegraphics[width=200pt]{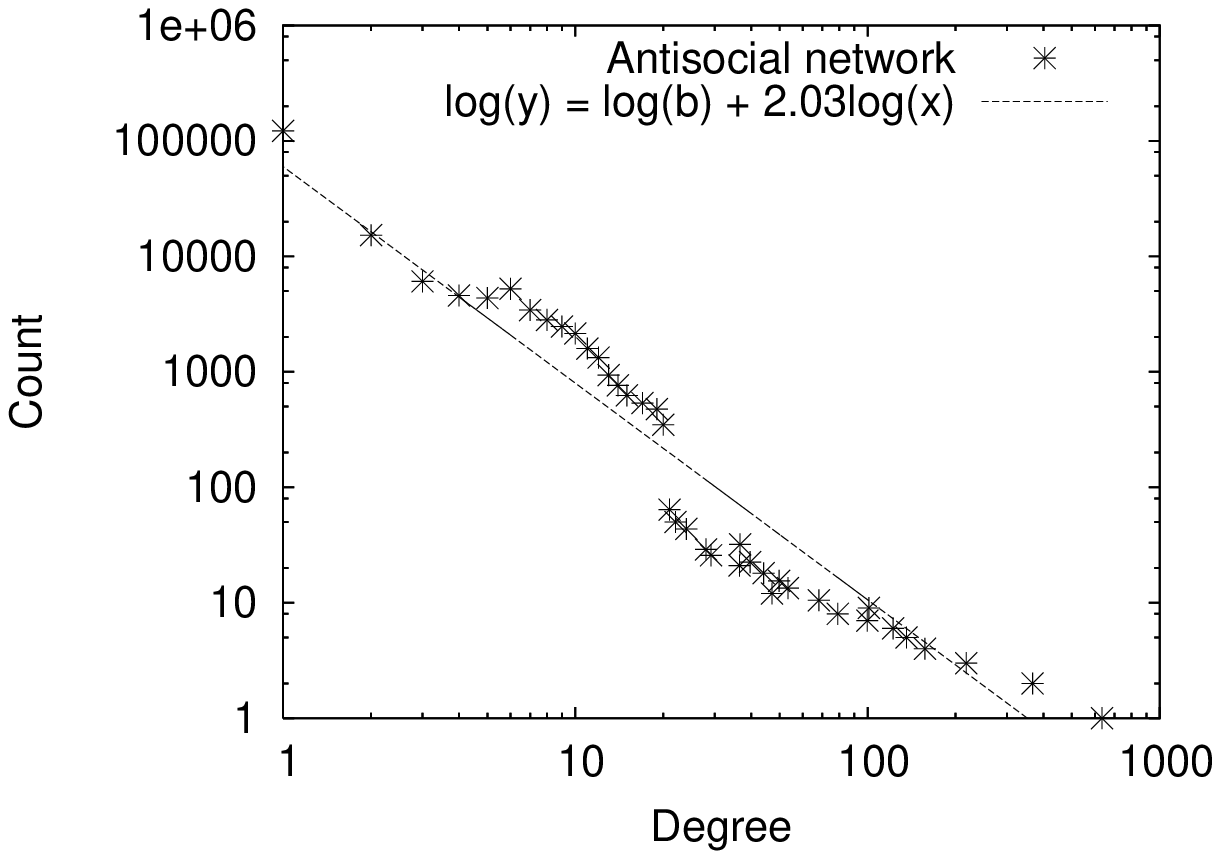}
 \caption{Average degree power-law distribution for social (top) and antisocial (bottom) networks.}
  \label{fig:dd}
 \end{figure}

Ebel, Mielsh and Bornholdt \cite{Ebel2002} analyzed a similarly constructed 
email network, although without drawing the distinction between spam and legitimate traffic. They 
characterized the degree $k$ distributions for the entire graph as a power law $P(k)\propto 1/k^\alpha$, 
with exponent $\alpha=1.81$. For the network composed exclusively of internal users they found a smaller 
exponent $\alpha=1.32$. Similarly we find power law degree distributions for the undirected versions 
of our four networks, with exponents $\alpha=1.82$ ($R^2=0.942$) 
for the full social network, $\alpha=2.03$ ($R^2=0.925$) for the entire antisocial network, see Figure~\ref{fig:dd}, 
and $\alpha=1.22$ ($R^2=0.958$) and $\alpha=1.79$ ($R^2=0.831$) for the internal social and 
antisocial networks, respectively. It is remarkable that our results are broadly consistent with 
those of \cite{Ebel2002}, for entirely different data. We find a tendency for the exponent 
to be larger for antisocial behavior, which suggests that the true social exponent may be 
over estimated if the two traffics are not separated. The lower values of $R^2$ for the antisocial 
networks suggest that the power law model is more adequate to represent social networks than their antisocial 
counterparts. Despite these differences, the degree distribution is a weak discriminator between social 
and antisocial behavior and is clearly affected by incomplete knowledge of parts of the network, 
which is a consideration whenever external users are included.  Such lack of knowledge results in the incorrect 
shift of external users to lower degree, and consequently leads to larger estimates of the exponent $\alpha$. 
Thus both the failure to exclude spam traffic and the incomplete knowledge of links between external users 
contribute to  overestimations of the exponent $\alpha$.

\begin{figure}[t!]
  \centering
\includegraphics[width=200pt]{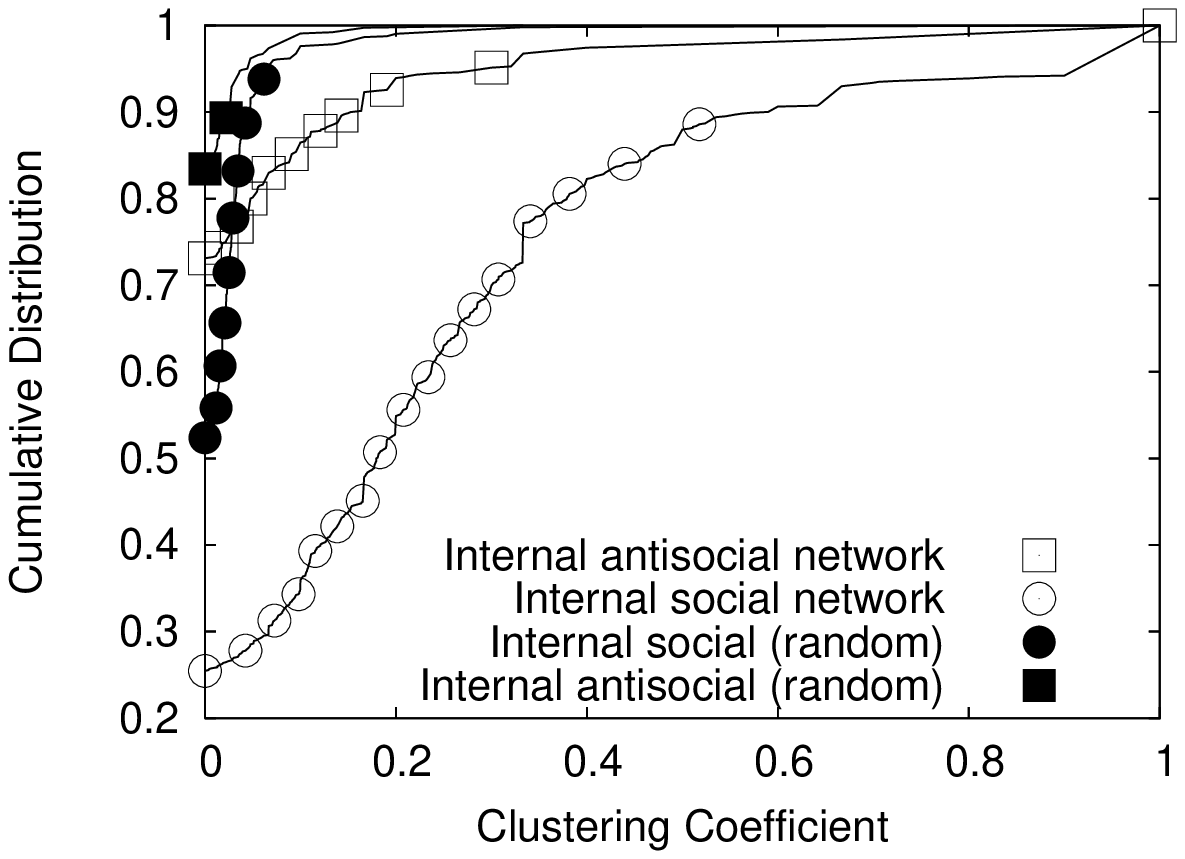}
\includegraphics[width=200pt]{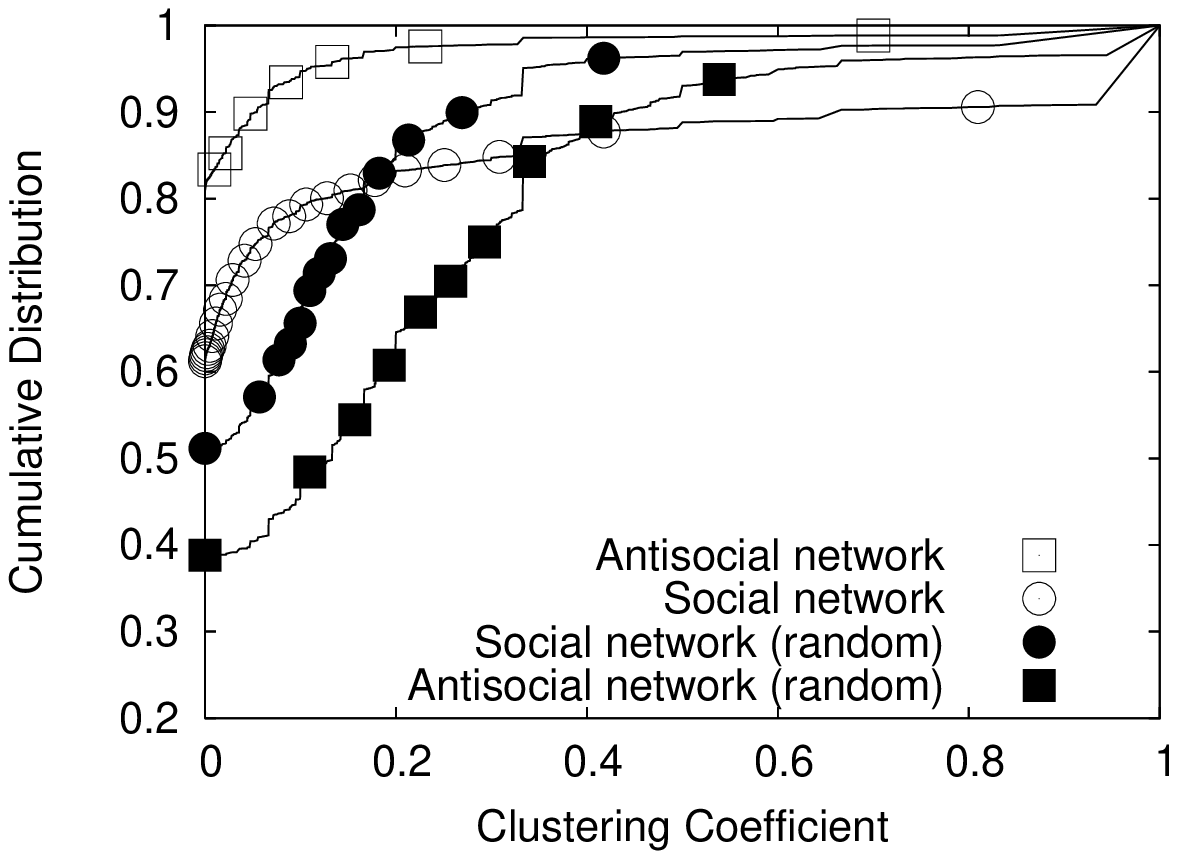}
 \caption{Distribution of the clustering coefficient for social, antisocial networks and their corresponding randomized networks with preserved degree sequence for 
the internal networks (top) and complete networks (bottom) 
built from messages exchanged between all users, internal and external.}
  \label{fig:cc}
 \end{figure}

Next, we recall that according to Newman and Park~\cite{Newman2003}, high clustering
coefficient and positive assortative mixing are two graph theoretical quantities 
typical of social networks. 
Therefore, we investigate whether these two structural properties of email graphs can distinguish
the social imprint of legitimate email communication from the antisocial characteristics of spam. In order
to do so we compare the average values of these network measures determined for networks constructed from actual data with corresponding values obtained for networks with randomized links, with the same degree sequence.

Indeed, considering the undirected versions of our networks, the average 
clustering coefficient over the internal social network is $C=0.241\pm0.008$, 
whereas the clustering coefficient in the internal antisocial network is much lower, at $C=0.052\pm0.006$. 
These results compare to the clustering coefficient of internal domain users of $C=0.154$, found by Ebel {\it et al.}~\cite{Ebel2002}. 
Considering the networks that include external users, whose neighbors are only known incompletely, 
we find $C=0.137\pm0.003$ for the social network and $C=0.026\pm0.001$ for the antisocial network, 
in contrast with a $C=0.003$ for the entire network of Ebel {\it et al}~\cite{Ebel2002}. Figure~\ref{fig:cc} shows the distribution of the clustering coefficient for social, antisocial and their corresponding random networks.

All four networks contain a significant fraction of their nodes with vanishing clustering coefficient, but this proportion is much higher for graphs that include external users and/or antisocial 
components. Specifically, $61\%$ of all nodes in the entire social network have $C=0$, while this becomes more than $81\%$ for the entire antisocial component. The internal social network has only $25\%$ of its nodes with $C=0$, compared to $73\%$ for the internal antisocial network. These features  indicate that there are clear 
differences on average between clustering in a social and an antisocial components of email networks, but also 
that low clustering is not a sufficient condition for a node to be associated with antisocial behavior.
Similarly to the analysis of the degree distribution these results also indicate that the separation of the two traffics is important in order to identify the truly social component. Failure to do so will result in the 
underestimation of the average social network transitivity.

We now analyze the nature of degree correlations between nodes by computing
the corresponding Pearson correlation coefficient~\cite{Newman2002b} $r$
\begin{eqnarray}
\label{eq:pearsoncoefficient}
r = \frac{ \sum_i{j_i k_i}  -  M^{-1} \sum_i{j_i} \sum_{i'}{k_{i'}}} {\sqrt{[\sum_i{j_{i}^{2}} - M^{-1}(\sum_i{j_ i})^2][\sum_i{k_{i}^{2}}  - M^{-1} (\sum_i{k_i})^2]}},
\end{eqnarray}
where $j_i$ and $k_i$ are the excess in-degree and out-degree of the vertices that the {\it i}th edge leads into and out of, 
respectively, and $M$ is the total number of edges in the graph.

The expectation of assortative mixing by degree in a social network of email is not obvious. 
In fact as we argued above, a user's degree is a very variable property, that can be easily changed 
drastically by the inclusion of the user's address in, or by the use of, distribution lists. 
This common use of email can create huge imbalances of degree between senders and 
recipients and may generate negative values for the Pearson coefficient 
even for groups of legitimate users.  If this can be expected of the degree correlation in the 
social network, then such an effect should be even more pronounced in the antisocial graph. 
There, spam senders follow the strategy of increasing their degree indiscriminately 
and maximally, and consequently reach on average a population of recipients with much lower degree, 
which are statistically much more abundant for a scale free degree distribution.

These qualitative expectations are borne out by estimation of $r$. Using (\ref{eq:pearsoncoefficient}) we computed the Pearson coefficient $r$ for each of the four {\it directed} networks, and obtained $r=-0.135$ for the entire social network (with  $r=-0.082$ for its corresponding randomized network), $r=-0.139$ ($-0.111$) for the entire antisocial network, and $r=0.232$ ($0.095$) and $r=0.049$ ($0.073$) for the social and antisocial internal networks, respectively.  Standard errors are smaller than $1\%$. Moreover, we observed that the positive value of $r$ for the internal social network is the result of an approximately linear correlation between the out degree of the sender and the in degree of the recipient. Such systematic correlation across degree is absent for the other three networks, with the difference that for networks containing external users there is an average imbalance between the degrees of senders and recipients that leads to a negative $r$.  As we can see from $r$ values, the social networks show significantly stronger assortativity (internal social network) and dissassortativity (social network) than their corresponding randomized networks.  On the other hand, there is a much less significant difference between the assortativity of real networks and their corresponding randomized versions in the antisocial case. 

We conjecture that the more negative Pearson coefficient for the complete
social network, which includes external users, is the result of the widespread subscription to legitimate distribution lists, such as those 
related to news, promotions, etc~\footnote{Recall that, unlike the internal social network, 
node degrees in our entire social network are not constrained, and thus, may represent 
distribution lists.} We verified to the extent possible, given that email user identifiers 
are made anonymous but domains are present, that external distribution 
lists are the main source of degree imbalance for the external social network.

%
%

In summary, we see that the consideration of this set of standard network measures places networks 
of email communication in a unique position. On the one hand, the legitimate component of a
 completely known email network shares its transitivity and positive degree correlation properties
 with other social networks. Unlike some other social networks however its degree distribution is scale 
free and characterized by a small exponent, which implies that, although the distribution 
remains normalizable, no finite moments exists as the network size goes to infinity ($2>\alpha>1$). 
This property is a direct result of the low cost of adding additional recipients to a message, and makes 
statistical estimation of degree correlations over email networks very sensitive and network size dependent, 
if not altogether ill defined.

In spite of these properties, the antisocial network built from the exchange of spam messages, 
has definite properties, showing negligible transitivity and assortative mixing near their corresponding random network with preserved degree sequence. 
Moreover, our analysis shows that, in contrast to previous expectations \cite{Newman2003}, 
social email networks involving users that are external to the local domain may present a 
negative degree correlation, presumably reflecting in part the incomplete knowledge of external links, 
but also resulting from message exchanges characteristic of email, such as the widespread subscription 
to legitimate distribution lists.

%
%

\begin{table*}
\centering
	\begin{tabular}{c c c c c } 
	
\hline										
Network 		& Internal	 social	& Internal  	& \multirow{2}{*}{Social} 	& \multirow{2}{*}{Antisocial}	\\ 
measure 	& social			& antisocial	&  						& 			 				\\ \hline		
Degree distribution ($\alpha$)		&$1.22$&$1.79$&$1.82$&$2.03$		\\ \hline
Clustering coefficient (real/random)& $0.2409/0.0188$ & $0.0521/0.0103$ & $0.1374/0.0089$ & $0.0261/0.0124$ 	\\ \hline
Assortative mixing (real/random)	&$0.2324/0.0946$&$0.0493/0.0727$&$-0.1347/-0.0824$&$-0.1387/-0.1110$	\\ \hline
Preferential exchange ($\langle E \rangle$)	& $0.27568$ & $0.06246$ & $0.03288$  & $0.00007$  			\\ \hline
	\end{tabular}
	\caption{Summary results for structural measures applied to the Social (legitimate email) and Antisocial (spam exchange) total networks and to those restricted to internal traffic within the domain.} 
\end{table*}

These differences suggest mechanisms to differentiate legitimate 
human collaboration from opportunistic behavior on the basis of network structure, 
and have indeed been proposed as the basis for spam detection algorithms 
\cite{Sruti,Boykin}. However, much remains unsatisfactory about the transitivity and assortative
mixing measures as means to characterize patterns of human communication. The most serious flaw is 
that their estimation relies on the knowledge of all neighbors of each node.  
This is not possible beyond a small subset, corresponding to users in the local
domain; a general problem of the construction of any network.  
A solution to this problem is the consideration  of quantities that characterize the dynamics of 
communication links between senders and recipients  
directly, without reference to third parties.  In other words, it is key to investigate whether the social and antisocial nature of a given node can be inferred from its dynamical behavior, 
even given incomplete knowledge of the social network of all its neighbors.  

\section{Temporal patterns of email communication}
\label{sec:temporal}

We start with the simplest measure of communication between two users: reciprocity~\cite{Garlaschelli}.  We build a simple coefficient of preferential exchange $E_i$ for user $i$ as: 
\begin{eqnarray}
E_i = 1 - \left\vert \frac{ \sum_{j \in C_i} \left[ k(j \rightarrow i) - k(i \rightarrow j) \right]  } { \sum_{j \in C_i} \left[ k(j \rightarrow i) + k(i \rightarrow j) \right] } \right\vert
\end{eqnarray}
where $C_i$ is the set of all users that have contact with user $i$ within a given time period, and $k(j\rightarrow i)$ is the number of messages sent by user $j$ to $i$. Therefore,  $0\leq E_i \leq 1$, with the lower end corresponding to no message being replied to, and the upper end to every 
message obtaining a response. This can be further averaged  
over all users to generate network expectation values $\langle E \rangle$. 
Considering internal as well as external users, we find $\langle E \rangle = 0.0329\pm0.0005$ in 
the social network, whereas a significantly lower $\langle E \rangle = 0.00007\pm0.00002$ is 
observed in the antisocial network. Values of $\langle E \rangle = 0.2757\pm0.0083$ and 
$\langle E \rangle = 0.0625\pm0.0056$ are found in the internal social and antisocial networks, 
respectively. Therefore, antisocial networks are naturally associated with small (but potentially non-zero)  reciprocity, whereas social networks, particularly those containing  legitimate 
users whose behavior we know completely, are associated with the highest 
reciprocity. 

Up to this point we concentrated on the structure of the network of interactions mediated by email messages. In its construction as a graph we have not paid attention to the detailed temporal structure of message exchanges.  An interesting question then is whether the dynamical properties of email traffic  can distinguish different types of social relations. This question has recently become a subject of interest. Eckmann, Moses and Sergi \cite{Eckmann2004} have shown that coherent structures emerge from the temporal correlations between time series 
expressing short periods of intense message exchange between groups of users. 
Barabasi \cite{Barabasi2005}, on the other hand, has shown that the distribution 
of time intervals between email messages sent by a single user may be well described by a power law distribution $P(\tau)\sim \tau^{-\gamma}$ with $\gamma \simeq 1$,  with bursts of activity  alternating with long silences. 

Both these characterizations identify properties of legitimate email traffic - temporal correlations between users 
and inter-message time statistics - that are thought to be exclusively social and thus not shared by the antisocial  traffic component. In fact intense email exchanges between small groups of users are to be expected
in patterns of human communication, creating the correlations observed by Eckmann, Moses and Sergi \cite{Eckmann2004}. Barabasi in turn suggests that the power law statistics he observed can be explained in terms of a queueing model which encodes prioritization of tasks driven by human decision making. 

\begin{figure}[t!]
  \centering
  \includegraphics[width=209pt]{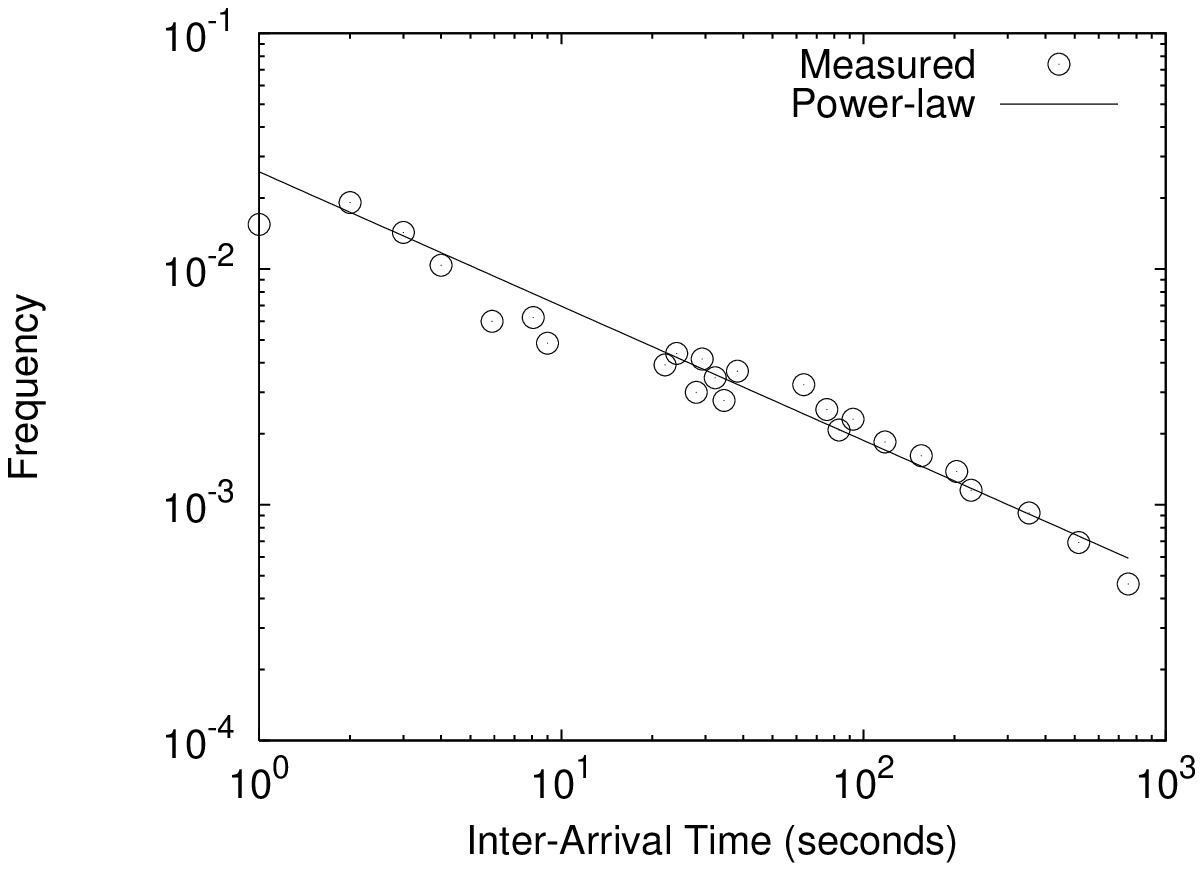}
  \includegraphics[width=200pt]{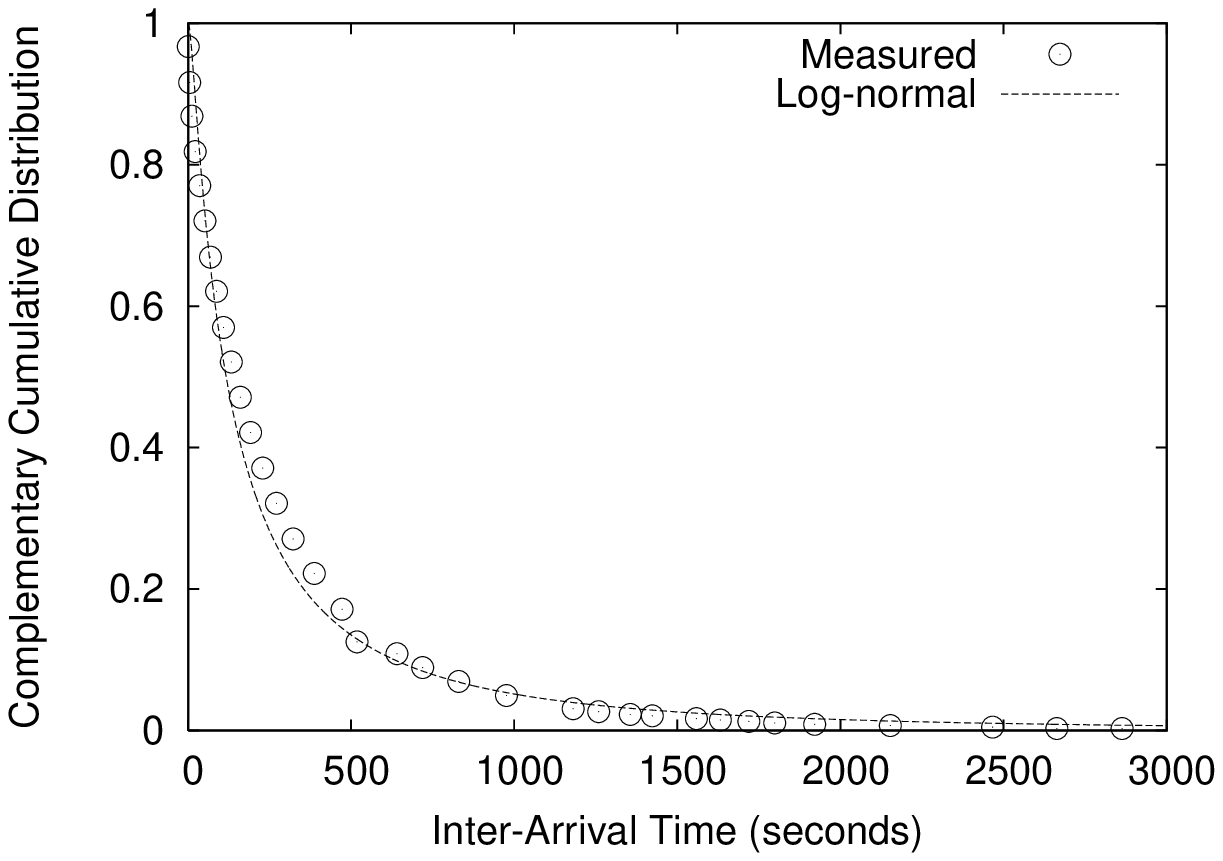}                                                                              
 \caption{The distribution of the intervals between sent messages for two of the most active senders of legitimate email. While figure \ref{fig:barabasi-distribution} (top) shows approximate power law statistics with power $\gamma=0.55$ ($R^2=0.96$), the distribution of figure~\ref{fig:barabasi-distribution} (bottom), for a different user, is better described by a log-normal distribution.}
\label{fig:barabasi-distribution}
 \end{figure}

Although suggestive, these interesting results were obtained for selected senders and receivers of email. 
Consequently it remains unclear whether they hold for the general user or for aggregated groups of users.  We have in fact attempted to verify Barabasi's findings in our log but obtained mixed results with some users showing the suggested power law behavior and others manifestly not, see Fig.~\ref{fig:barabasi-distribution}. Similar results were reported in Ref.~\cite{Stouffer}. 

\begin{figure}[t!]
  \centering
\includegraphics[width=209pt]{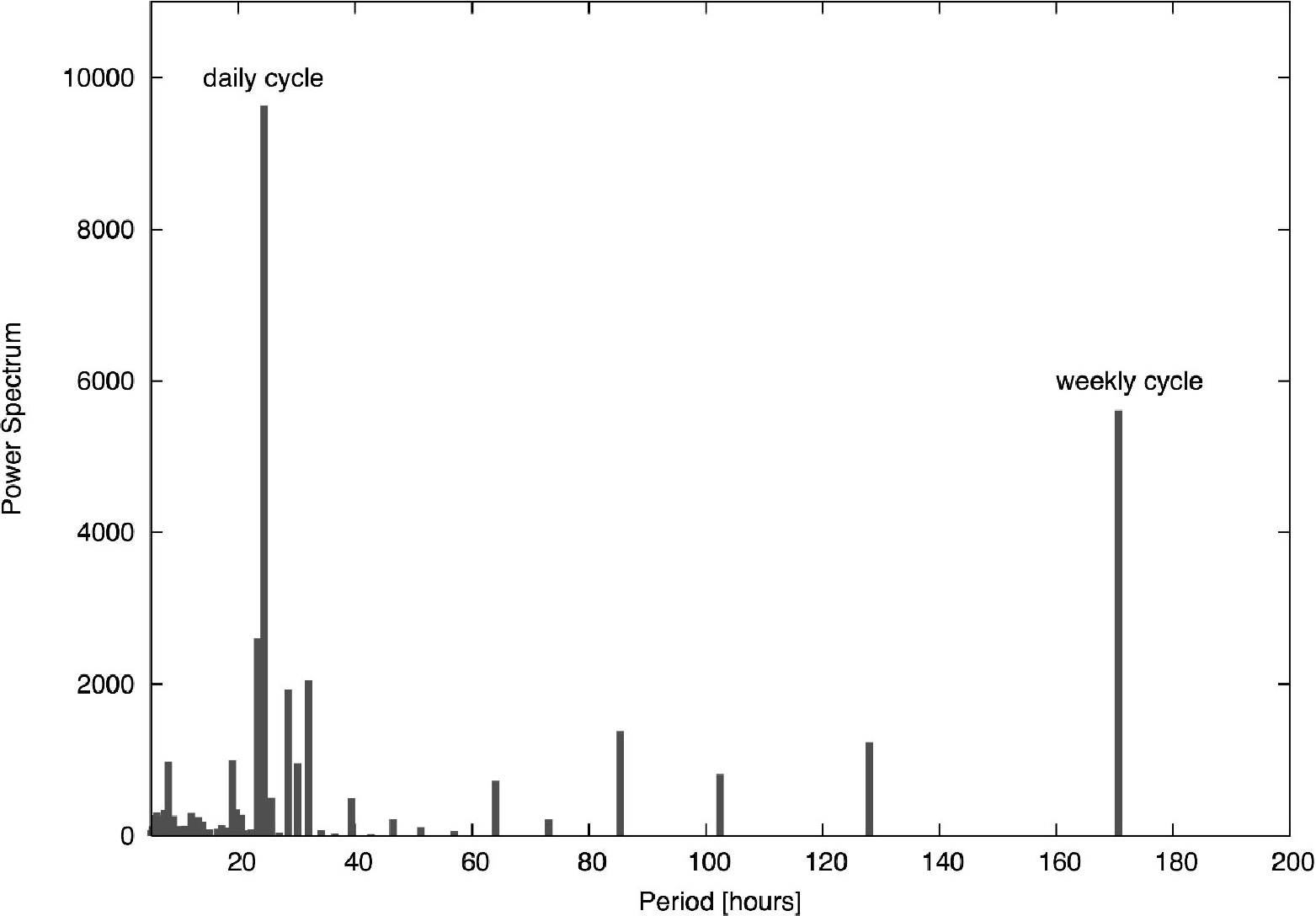}
\includegraphics[width=200pt]{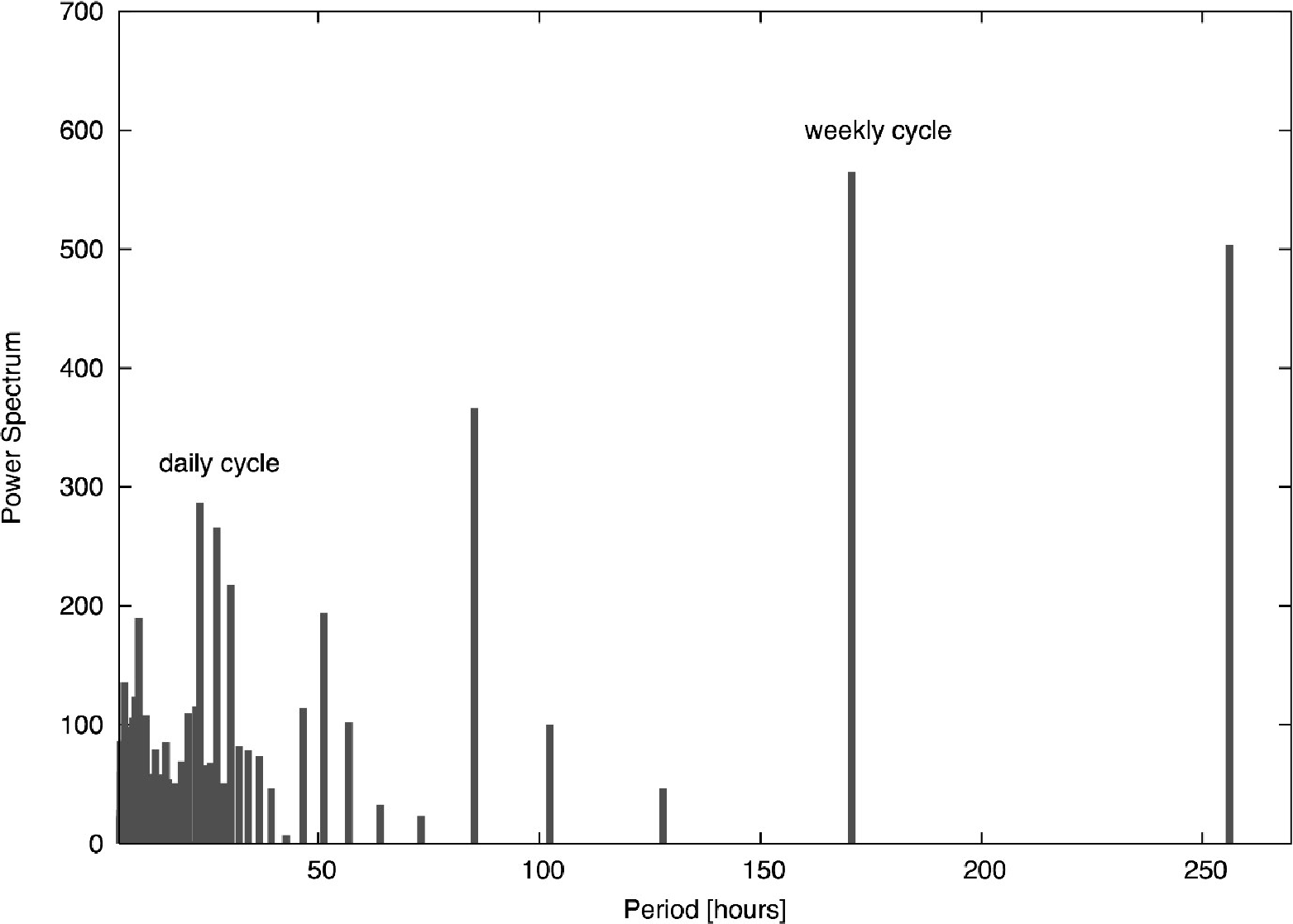}
 \caption{Temporal power spectrum of the legitimate (top) and spam (bottom) traffics aggregated over all users in each class. Daily and weekly periods are features of both traffics,  but dominate legitimate email exchange. The power spectrum of spam traffic is more uniform at short times.}
\label{fig:powerspectra}
 \end{figure}

To evade effects of variability associated with individual users, we chose to investigated the statistics of our social and antisocial aggregate traffics through averaging over the behavior of all users in each class.  The first obvious temporal property of email traffic is its non stationarity, see Fig.~\ref{fig:powerspectra}.  This feature creates  difficulties for any attempt at statistical estimation. Social email traffic in particular shows large temporal variations, from night to day, working days to weekends, and for our data set, strong seasonality associated with the  academic calendar. Antisocial traffic displays weaker non-stationarity, see Fig.~\ref{fig:powerspectra}.

The second temporal feature of email traffic is an immediate result of the power law degree distributions 
described above. The majority of users do not communicate often with many others, but have instead low degree 
associated with an infrequent and often irregular usage of email. This means that the typical email user in 
our data - and, we believe, in most other large email networks - does not show time coherence 
with others, nor is he/she necessarily under the constraints of temporal optimization suggested by Barabasi. 

To circumvent some of these difficulties, we attempted to identify statistical temporal patterns of communication  that are characteristic of the social vs. antisocial aggregated traffics. In so doing we average over the behaviors 
of many users.  Specifically, we represent temporal patterns of message arrival through the definition of a state in terms of a communication {\it word} of size $L$.  The dimension $L$ is the number of time intervals, or letters, in the communication word, which is written as a vector $ W = [i_1, i_2, \ldots, i_L] $. 
The simplest representation of the traffic is through a binary assignment, where the value  of $ i_j $ is set  to 1 if one or more messages were exchanged in the corresponding time interval,  or $ i_j=0 $ otherwise. i.e. 
\begin{eqnarray}
W= \left[ 0 1 001\ldots 01\right],
\end{eqnarray}
where there are $L$ boolean variables, each corresponding to the exchange, or not, of a message in consecutive time periods $\Delta t$. For stationary processes the probability of a message exchange occurs with a fixed probability per unit time. The representation of time series in terms of binary words is familiar from other contexts in physics and information theory~\cite{Bialek,Crutchfield}, from the analysis of the time evolution of dynamical systems, to trains of action potential in neuronal activity \cite{Spikes} or bit streams in noisy communication channels. The entropy of the distribution and its variation with the word size $L$ give us in fact some of the essential properties of the dynamical rules that generate these dynamical patterns \cite{Bialek,Crutchfield}.

To illustrate these statements consider the simplest statistical model that  generates a binary time series subject to a given message arrival rate $p$. Then $p$ can be written as the probability to obtain a 1 at each letter. If we further assume that bits corresponding to different letters are uncorrelated then the bit value at each letter can be regarded as the result of an independent Bernoulli trial. 

Under these assumptions the probability of a given number of events $k$ in $L$ trials (bins) 
 is well known to be given by the binomial distribution
 \begin{eqnarray}
 f(k;L,p)=\left(^L_k \right)~ p^k~(1-p)^{L-k}.
 \end{eqnarray}
Moreover the probability of a sequence with the same number of events is the same regardless of their order, as each occurrence is independent for different bins. Thus to obtain the probability for a particular sequence of $k$ events in $L$ bins we must divide by the number of possible arrangements $\left(^L_k \right)$. Then the probability for a particular sequence or binary word with $k$ ones and length $L$ is
\begin{eqnarray}
p_W= \frac{p^k ~(1-p)^{L-k}}{\sum_{k'=0}^L ~ \left(^L_{k'} \right)~ p^{k'}~(1-p)^{L-{k'}}} =p^k ~(1-p)^{L-k} . 
\end{eqnarray}
Because all words with a given number $n$ of 1s are equally likely, their 
probability is $p_W(n;L,p)=p^n~(1-p)^{L-n}$. This implies that the Shannon entropy of the time series can be written as 
\begin{small}
\begin{eqnarray}
H &=& -\sum_W p_W \log_2 (p_W)  = -\langle k \rangle \log_2 \left( \frac{p}{1-p} \right) - L\log_2 (1-p) \nonumber \\ &=&m~L, 
\end{eqnarray}
\end{small}
with $\langle k \rangle = L p$. Thus, in the absence of temporal correlations, the Shannon entropy is a strict linearly growing function of the word length $L$, with slope $m = - (1-p) \log_2 (1-p)   - p \log_2 p > 0$. 

These expressions become especially simple if the temporal bin for each letter is chosen such that $p=1/2$, in which case $m=1$ is maximal. This independent message model (IMM) is the maximal entropy distribution for a traffic characterized by an average message arrival probability $p$. Real traffics, which show temporal structure, must therefore display lower entropy  relative to the idealized IMM message stream. We refer to the difference of the traffic entropy to that of the corresponding $H_{IMM}(L)$, measured with the same average choice of $p$, as the traffic's structural information, for a given $L$.    

\begin{figure}[th!]
\centering
	\includegraphics[width=200pt]{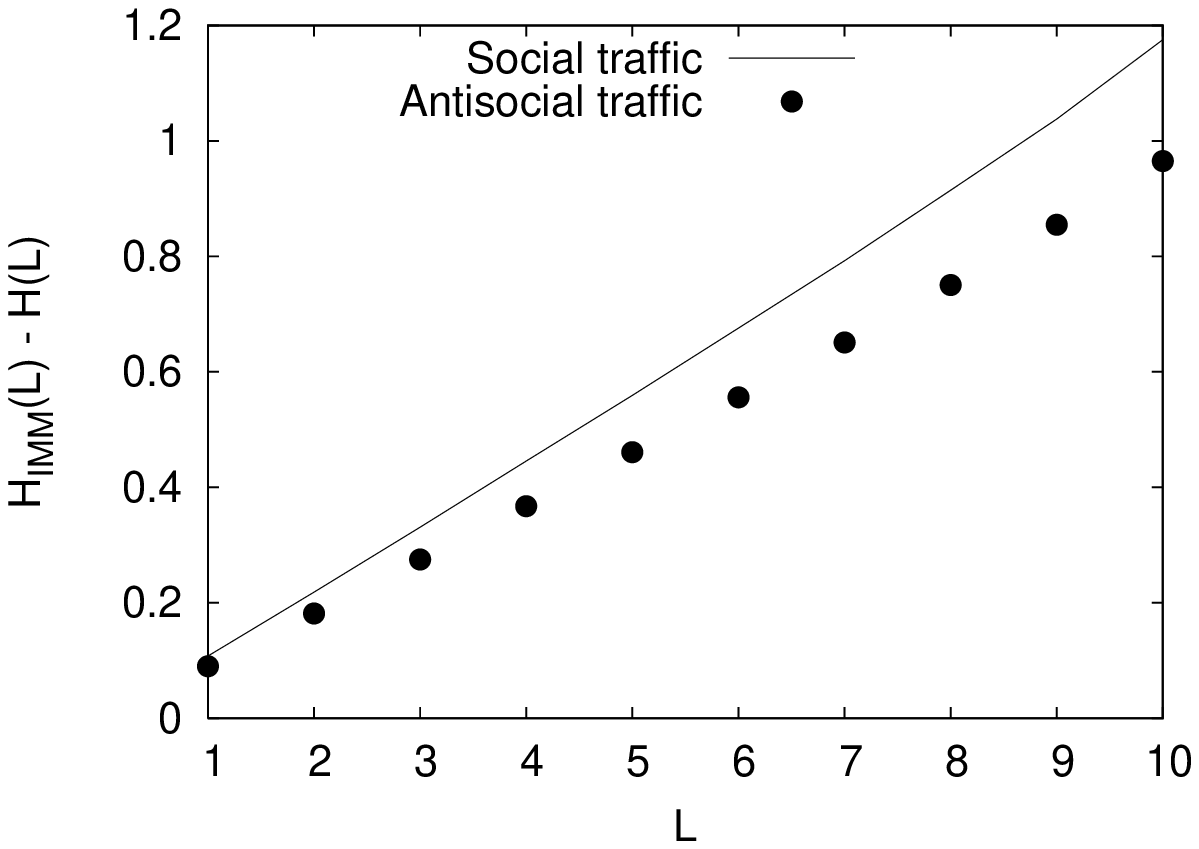}
	\includegraphics[width=200pt]{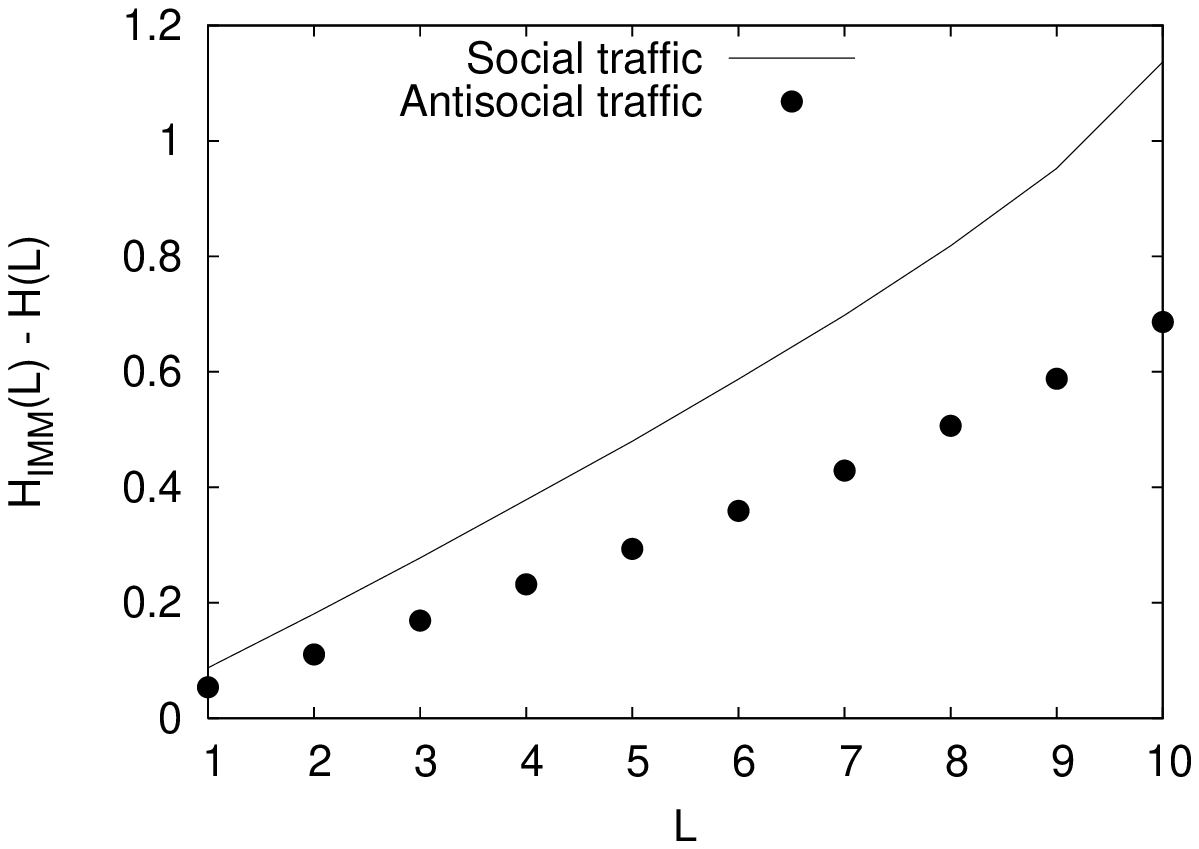}
\caption{The variation of the difference between the independent message model entropy $H_{IMM}(L)$ and the entropy of the legitimate and spam traffics $H(L)$, with word size $L$,  during work (top) and non-work (bottom) periods. All word probability distributions were constructed by normalizing the time bin for each letter word so that $p=1/2$. As a result the time bin for each letter of the social traffic during work hours was set to 4s, and 11s for the corresponding non-work period. Time bins for the antisocial traffic were set at 4s during work hours and 5s  otherwise. The slight excess curvature for large $L$ is the result of poorer estimation of rare long words.}
\label{fig:entropy}
\end{figure}
 
Figure~\ref{fig:entropy} shows the difference between the entropy of the independent message model  and the real traffics, legitimate and spam. We aggregated the data into two temporal periods: 
work hours (i. e. the period from 8AM to 8PM of the weekdays, except holidays, in the log) and remaining times  which we refer to as non-work hours. 

The results show that the social email traffic has lower entropy (higher structural information) than the antisocial traffic for both work and non-work periods. This difference becomes more noticeable the larger the word, thus capturing longer patterns of communication and the presence of time correlations. 
The difference between the independent message model, where for $p=1/2$ all words are equally likely,  and the real traffics is that in the latter words with many 1s (0s) are suppressed while the probability of words with two to three 1s separated by one to three 0s is enhanced. The difference between social and antisocial traffics is more subtle, with social email traffic displaying a greater probability for words with an isolated message in a long stream of silence. These structures are reminiscent of those found by Barabasi \cite{Barabasi2005}, but display less definitive statistical signatures. Nevertheless, we see that both social and antisocial traffics are far from random, and that social email shows stronger temporal structure with a high probability for long silences and bursts of a few messages.

%
%

\section{Discussion and Conclusions}
\label{sec:conclusions}

We have shown that the richness of behaviors in human communication - both symbiotic and opportunistic or antisocial - is present in the structure of networks of email communication and can be quantified via graph theoretical and time series analysis. Opportunistic nodes display antisocial behavior that can be captured graphically through the absence of definite metrics present in other social networks. Perhaps  even more directly, antisocial email traffic can be identified by a greater statistical simplicity (higher entropy) in temporal patterns of communication, typical of the fact that each sender/recipient relationship is not developed to be unique and the same schemes are used to reach many recipients indiscriminately. Moreover, the ease to exchange email messages that leads to these opportunistic behaviors also has consequences for the truly social component of the network, which exhibits a power law degree distribution with a small exponent and, in some cases, small or negative assortative mixing by degree. We believe that the quantitative characteristics of antisocial communication patterns observed here for email networks are probably general to other opportunistic social behaviors, bound to be present in other networks of human interaction. 

\begin{acknowledgments}
The authors thank CNPq (Brazilian Council for Research and Development), 
Los Alamos National Laboratory and the Central Bank of Brazil for support.
\end{acknowledgments}

\end{document}